\documentstyle[aps,preprint]{revtex}
\begin{document}
\title{Mobility of Bloch Walls via the Collective Coordinate Method}
\author{M.A. Desp\'{o}sito\thanks{E-mail: mad@df.uba.ar}$^{1}$, 
A. Villares Ferrer$^{2}$, A.O. Caldeira$^{2}$ \\
and A.H. Castro Neto$^{3}$}
\address{$^{1}$Departamento de F\'{\i}sica, Facultad de Ciencias Exactas y Naturales,\\
Universidad de Buenos Aires, RA-1428, Buenos Aires, Argentina.}
\address{$^{2}$Instituto de F\'{\i}sica ``Gleb Wataghin'',\\
Departamento de F\'{\i}sica do Estado S\'{o}lido e Ci\^{e}ncia dos Materiais,
\\
Universidade Estadual de Campinas, 13083-970, Campinas, SP, Brasil.}
\address{$^{3}$Dept. of Physics, University of California,
Riverside CA, 92521, USA.}
\maketitle

\begin{abstract}
We have studied the problem of the dissipative motion of Bloch walls
considering a totally anisotropic  one dimensional spin chain in the presence
of a magnetic field. Using the so-called ``collective coordinate method'' we
construct an effective Hamiltonian for the Bloch wall coupled to the magnetic
excitations of the system. It allows us to analyze the Brownian motion of the wall in
terms of the reflection coefficient of the effective potential felt by the
excitations due to the existence of the wall. We find that for finite
values of the external field the wall mobility is also finite. The spectrum
of the potential at large fields is investigated and the dependence of the
damping constant on temperature is evaluated.
As a result we find the temperature and magnetic field  
dependence of the wall mobility.
\end{abstract}

\pacs{?}

\section{Introduction}

It is a well-known fact that ultimately due to magnetic dipole interaction,
different domains are formed in magnetic systems \cite{Lan}. In many
situations, the physical region separating two different magnetic
domains---the domain wall---must be treated as a physical entity because it
has a characteristic behaviour when acted by external agents. For instance,
it is known that the response of a magnetic system to a frequency dependent
external magnetic field depends on whether domain walls are present \cite{Dillon}. 
Domain walls can also move throughout the system and this motion happens to be 
dissipative \cite{Land}.

A particularly interesting kind of domain wall is commonly found in low
dimensional ferromagnetic systems. These are the so-called Bloch walls\cite
{Malozemoff}. It is known that these walls perform dissipative motion \cite
{Land} due to the presence of the elementary excitations which can be
scattered by the wall as it moves and the momentum transferred to them
reduces the speed of the wall.

The primary aim of this work is to study the influence of  finite
temperatures in the mobility of these Bloch walls. For this purpose we start
by considering a microscopic model for a one dimensional ferromagnet
containing hard and easy-axis anisotropies and subject to an external
magnetic field. A semiclassical picture provides us with the localized solutions
for the spin configurations which are the solitons corresponding to the walls.

Making use of a recently developed method for the analysis of the
dissipative dynamics of solitons \cite{CN,Caldeira}, in which the
``collective coordinate method'' \cite{R.R.} is used to transform the
original Hamiltonian into  one of a particle coupled to an infinite set of
modes, we show that the Bloch wall behaves like a Brownian particle. The
advantage of using this method is that we keep  closer contact with the
microscopic details of the system and the mobility is naturally calculated
as a function of the temperature. The information from the microscopic
scattering processes between the Bloch wall and the residual modes can be
obtained from the knowledge of the phase shifts of the associated spectral
problem. In the case of reflectionless potentials, as it happens for
vanishing anisotropies or external field, the motion of the wall is undamped. 
If this is
not the case, the reflection coefficient does not vanish and the mobility is
finite.

The outline of this paper is as follows. In Sec. II we present the model.
The dynamics of its static solution is investigated in Sec. III and there we also 
show how to obtain an effective Hamiltonian for the Bloch wall coupled to the
residual magnetic excitations. In Sec. IV the mobility of the Bloch wall is
studied in terms of the scattering phase shifts of the second variation
problem. The case of large external fields is investigated in Sec.V where
the phase shifts and the damping constant are explicitly evaluated. Finally,
we present our conclusions in Sec. VI.

\section{The Model and its static solutions}

In this work we consider a one dimensional magnetic system composed by an
array of spins lying along the $\hat{z}$-direction. Furthermore, let us
assume that there is an easy-plane anisotropy which tries to keep the spins
on the $x$-$y$ plane and, on top of this, an in-plane anisotropy tending to
align them along the $\hat{x}$-direction. This is a totally anisotropic model 
which is described by a XYZ model of magnetic systems defined by the
Hamiltonian

\begin{equation}
H=-\sum_{\langle ij\rangle }\left(
J_xS_i^{(x)}S_j^{(x)}+J_yS_i^{(y)}S_j^{(y)}+J_zS_i^{(z)}S_j^{(z)}\right)
-\frac \mu \hbar B\sum_iS_i^{(x)}  \label{Hspin}
\end{equation}
where ${J_x>J_y>J_z>0}$, ${S_j^{(\alpha )}}$ is the $\alpha $ component (${%
\alpha =x,y,z}$) of the $i^{th}$ spin of the system, ${\mu }$ is the modulus
of the magnetic moment on each site and $B$ is the external magnetic field.
The ferromagnetic $XYZ$ model is actually defined for ${B=0}$ and this is
the starting point of our analysis. As we can see from ($\ref{Hspin}$), the
ground state of this system is the configuration where all the spins are
aligned in the $\hat{x}$-direction. However, there is another possible
configuration which is a local minimum of the energy functional and cannot
be obtained from the previous uniform configuration by any finite energy
operation.

Let us imagine that we describe our spins classically by vectors

\begin{equation}
{\bf S}_i=S(\sin \theta _i\cos \varphi _i,\sin \theta _i\sin \varphi _i,\cos
\theta _i)  \label{semispin}
\end{equation}
where $\theta _i$ and $\varphi _i$ are the polar angles of the $i^{th}$
spin. In this representation the above-mentioned configuration consists of
all $\theta _i$'s equal to ${\pi /2}$ \ and \ $\varphi _i$'s equal to zero
or $\pi $. However, there are other configurations in which $\theta _i=\pi
/2,\varphi _i=0$ if ${i\rightarrow -\infty }$ and ${\varphi _i=\pi }$ if ${%
i\rightarrow \infty }$ which are approximately (only because $\theta _i$'s
may slightly vary \cite{Stamp}) local minima of the energy functional of the
system. So, ${\bf S}_i$ winds around the $\hat{z}$ direction starting at ${%
(\theta ,\varphi )}$ = ${(\pi /2,0)}$ and ending at ${(\theta ,\varphi )}$ =
${(\pi /2,\pi )}$. The so-called $\pi $-Bloch wall \cite{Winter} is one
example of these configurations where $\varphi _i$ varies from $0$ to $\pi $
without making a complete turn around the $\hat{z}$ axis. Later on we will
see the specific form of this configuration when we consider the system in
the continuum limit. It will then be shown that Bloch walls are related to 
soliton-like solutions of the non-linear equations which control the spin
dynamics in the semi-classical approximation.

If we now turn the external field $B$ on what happens is that the degeneracy
between ${\varphi =0}$ and ${\varphi =\pi }$ is broken. For ${B>0}$ it is
clear from ($\ref{Hspin}$) that ${\varphi =0}$ has lower energy than ${%
\varphi =\pi }$ which is now a metastable configuration of the system. In
this circumstance the system still presents a local minimum of the energy
functional. The only difference is that whereas $\theta_i$ is still
approximately ${\pi /2}$, $\varphi _i$ starts and ends at zero as -${\infty
<i<\infty }$. The 2$\pi $-Bloch wall is now the configuration where $\varphi
_i$ winds only once around $\hat{z}$.

In any of the two cases mentioned above, there is no way we could spend a
finite energy to transform the Bloch wall into the uniform configuration. We
would need to turn an infinite number of spins over an anisotropy energy
barrier. We say that these two configurations are topologically distinct.

Another important point is that ($\ref{Hspin}$) is translation invariant and
this is reflected by the translation invariance of the Bloch wall. This
means that the region about which the spins wind up can be centered anywhere
on the $\hat{z}$ axis. In reality they can even move with constant speed
along that direction.

These structures can be obtained by mapping the original Hamiltonian ($\ref
{Hspin}$) into a $1+1$ field theoretical model such as the $\varphi ^4$,
sine-Gordon or any other appropriate model. This can be done by simply proceeding a bit
further with the semi-classical description for the spins we have seen in (%
\ref{semispin}) and writing the Hamilton equations of motion for $\theta
_i(t)$ and $\varphi _i(t)$. After having done that we take the continuum
limit $\theta _i\rightarrow \theta (z,t)$ and $\varphi _i(t)\rightarrow
\varphi (z,t)$ and write

\begin{equation}
\theta (z,t)\approx \frac \pi 2+\alpha (z,t),
\end{equation}
where ${\alpha (z,t)\ll \ 1}$. Assuming that the variations of $\varphi $ and
$\alpha $ from site to site of the spin chain are small and linearizing the
equations of motion with respect to $\alpha $ one obtains

\begin{eqnarray}
\dot{\varphi} &=&\alpha 2S(J_x\cos ^2\varphi +J_y\sin ^2\varphi -J_z) \\
\dot{\alpha} &=&a^2S(J_x\sin ^2\varphi +J_y\cos ^2\varphi )\frac{\partial
^2\varphi }{\partial z^2}-\sin \varphi \left[\frac{\mu B}\hbar +2S(J_x-J_y)\cos \varphi
\right]
\end{eqnarray}
where $a$ is the lattice spacing. Then, eliminating $\alpha $ from these
equations, we get an effective equation of motion for ${\varphi (z,t)}$ of
the form

\begin{equation}
\frac 1{c^2}\frac{\partial ^2\varphi }{\partial t^2}-\frac{\partial
^2\varphi }{\partial z^2}=-A_1\sin \varphi -A_2\sin 2\varphi  \label{mofi}
\end{equation}
where

\begin{equation}
c^2\cong 2a^2S^2J_xJ_y\left( 1-\frac{J_z}{J_y}\right) ,  \label{c}
\end{equation}

\begin{equation}
A_1=\frac {\mu B}{a^2SJ_x \hbar} ,  \label{A1}
\end{equation}
and

\begin{equation}
A_2=\frac 1{a^2}\left( 1-\frac{J_y}{J_x}\right) .  \label{A2}
\end{equation}

Notice that if $J_x=J_y$ one has $A_2=0$ whereas if $B=0$ it turns out that $%
A_1=0$. So, as we can derive the r.h.s. of (\ref{mofi}) from a potential
energy density $U(\varphi )$ given by

\begin{equation}
U(\varphi )=A_1(1-\cos \varphi )+\frac{A_2}2(1-\cos 2\varphi )  \label{pot}
\end{equation}
we see that $A_1$ controls the potential energy barrier due to the presence
of $B\neq 0$ and $A_2$ controls the anisotropy energy barrier.

The static solutions (${\partial \varphi /\partial t=0}$) of Eq. (\ref{mofi}%
) are obtained using that \cite{R.R.}

\begin{equation}
z-z_0=\int_{\varphi (z_0)}^{\varphi (z)}\frac{d\varphi ^{\prime }}{\sqrt{%
2U(\varphi ^{\prime })}}
\end{equation}
are the solitons of the system. In particular, the examples of Bloch walls
we gave above are the solitons

\begin{eqnarray}
\varphi (z) &=&2\tan ^{-1}[\exp \sqrt{2A_2}(z-z_0)]\quad \text{ if }A_1=0
\label{fia10} \\
\varphi (z) &=&4\tan ^{-1}[\exp \sqrt{A_1}(z-z_0)]\quad \;\text{ if }A_2=0
\label{fia20}
\end{eqnarray}
while for the general case of finite anisotropy and magnetic field the
solution is the 2$\pi $-Bloch wall

\begin{equation}
\varphi (z)=2\tan ^{-1}[\frac{\cosh \rho }{\sinh (z/\lambda )}]
\label{figen}
\end{equation}
where we define

\begin{eqnarray}
\lambda &=&1/\sqrt{A_1+2A_2} \\
\cosh \rho &=&\sqrt{1+\frac{2A_2}{A_1}}
\end{eqnarray}

The soliton (\ref{figen}) can be expressed as a superposition of two twisted
$\pi $-Bloch walls \cite{Braun} with arguments $\lambda ^{-1}(z-z_0)\pm \rho $.

We mention here there is another static solution of (\ref{mofi}), the so-
called {\it nucleus} \cite{Long} which correspond to a superposition of two
untwisted $\pi $-Bloch walls \cite{Braun}. This solution is topologically
distinct from the previous one.

\section{Dynamics of Bloch walls}

The quantum dynamics of our spin system can be analyzed by studying the
quantum mechanics of the field theory described by the action

\begin{equation}
S[\varphi ]=JS^{2}a\int_{-\infty }^{+\infty }\int_0^tdz\,dt\,\left\{ \frac 1{2c^2}%
\left( \frac{\partial \varphi }{\partial t}\right) ^2-\ \frac 12\left( \frac{%
\partial \varphi }{\partial z}\right) ^2-U(\varphi )\right\} . \label{action}
\end{equation}

The next step is to quantize the system described by (\ref{action}). The
standard way to carry this program forward is to evaluate \cite{R.R.}

\begin{equation}
G(t)=\text{tr}\int {\cal D}\varphi \ \exp \ \frac i\hbar S[\varphi ]
\label{gt}
\end{equation}
where the functional integral has the same initial and final configurations
and tr means to evaluate it over all such configurations.

As the functional integral in (\ref{gt}) is impossible to be evaluated for a
potential energy density as in (\ref{pot}) we must choose an approximation
to do it. Since we are already considering large spins ${(S\gg \ \hbar /2)}$%
, and consequently in the semi-classical limit, let us take this
approximation as the appropriate one for our case.

The semi-classical limit ($\hbar \rightarrow 0$) turns out to be very easily
tractable within the functional integral formulation of quantum mechanics
\cite{R.R.}. It is simply the stationary phase method applied to (\ref{gt}).
Moreover, since we are only interested in static solutions, the functional
derivative of $S$ happens to be the equation of motion (\ref{mofi}) when ${%
\partial \varphi /\partial t=0}$. Its solutions can be either constant
(uniform magnetization) or the solitons (Bloch walls) we mentioned in (\ref
{fia10}-\ref{figen}). Since we are interested in studying the magnetic
system in the presence of walls it is obvious that we must pick up one of
those localized solutions as the stationary ``point'' in the configuration
space and the second functional derivative of (\ref{pot}) should be
evaluated at this configuration.

When this is done we are left with an eigenvalue problem that reads

\begin{equation}
\left\{ -\frac{d^2}{dz^2}+U^{\prime \prime }(\varphi _s)\right\} \psi
_n(z-z_0)=\kappa _n^2\psi _n(z-z_0)  \label{Sle}
\end{equation}
where $\varphi _s$ is denoting the soliton-like solution about which we are
expanding $\varphi (z,t)$.

Now one can easily show that $d\varphi _s/dz$ is a solution of (\ref{Sle}) with $%
\kappa _n=0$. The existence of this mode is related to the translation
invariance of the Lagrangian in (\ref{action}) and this makes the functional
integral in (\ref{gt}) blow up in the semi-classical limit (Gaussian
approximation).

The way out of this problem is the so-called collective coordinate method which
was developed by field-theorists in the seventies (see \cite{R.R.} and
references therein). It consists of keeping the expansion of the field
configurations about $\varphi _s(z)$ as

\begin{equation}
\varphi (z,t)=\varphi _s(z-z_0)+\sum_{n=1}^\infty c_n\psi _n\left(
z-z_0\right)  \label{expfi}
\end{equation}
but regarding the $c$-number $z_0$ as a position operator. Eq. (11) is then
substituted in the Hamiltonian

\begin{equation}
H=JS^{2}a\int dx\left\{ \frac{c^{2}\Pi ^2}2+\ \frac 12 \left( \frac {d\varphi }{dz}\right) 
^{2}+U(\varphi )\right\} ,
\end{equation}
where ${\ \Pi =\frac {1}{c} (\frac{\partial \varphi }{\partial t})}$, which can be
transformed into \cite{Castro}

\begin{equation}
H=\frac 1{2M_s}\ (P-\sum_{mn}\hbar g_{mn}b{_n^{+}}b_m)^2+\sum \hbar \Omega
_nb{_n^{+}}b_n.  \label{Heff}
\end{equation}
where $\Omega _{n}\equiv c\kappa _{n}$.

In the Hamiltonian (\ref{Heff}), $P$ stands for the momentum canonically
conjugated to $z_0$,

\begin{equation}
M_s=\frac {2JS^{2}a}{c^{2}}\int_{-\infty }^{+\infty }dzU(\varphi _s(z))  \label{sm}
\end{equation}
is the soliton mass \cite{R.R.} and the coupling constants $g_{mn}$ are
given by

\begin{equation}
g_{mn}=\frac 1{2ia}\left[ \left( \frac{\Omega _m}{\Omega _n}\right)
^{1/2}+\left( \frac{\Omega _n}{\Omega _m}\right) ^{1/2}\right] \int dz\psi
_m(z)\frac{d\psi _n(z)}{dz}.  \label{gmn}
\end{equation}

The operators $b^{+}$ and $b$ are respectively creation and annihilation
operators for the excitations of the magnetic systems (magnons) in the
presence of the wall.

It should also be stressed that Eq.~(\ref{Heff}) is not an exact result. It
is only valid in the limit $\hbar \rightarrow 0$ or, to be more precise,
when $g^2\hbar \rightarrow 0$ where $g^2 \equiv \frac{1}{JS^{2}a}$ is the coupling 
constant that originally appears in $U(g,\varphi )$. It must also
be emphasized that we have neglected inelastic terms such as $b^{+}b^{+}$ or
$bb$ because these are only important if the wall moves at high speed $(v>c)$
originating Cherenkov-like radiation of the elementary excitations of the
medium. This approximation also means that the number of excitations in the medium is 
conserved.

\section{Mobility of the Bloch wall}

At this point we are ready to start to study properties such as the mobility
of the wall because we have been able to map that problem into the
Hamiltonian (\ref{Heff}) which on its turn has been recently used to study
the mobility of polarons, heavy particles and solitons in general. We shall
not discuss this specific problem in this paper and urge those interested in
the details of this calculation to follow them in references \cite
{CN,Castro,Neto,CyN,Des}.

The result that can be obtained reads \cite{CyN}

\begin{equation}
\gamma (t)=\frac \hbar {2M}\int_0^\infty \!\int_0^\infty d\omega d\omega
^{\prime }S(\omega ,\omega ^{\prime })(\omega -\omega ^{\prime })[n(\omega
)-n(\omega ^{\prime })]\cos (\omega -\omega ^{\prime })t  \label{g}
\end{equation}
where $\gamma (t)$ is the damping function (the inverse of the mobility),
\begin{equation}
n(\omega )=\frac 1{e^{\beta \hbar \omega }-1}
\end{equation}
is the Bose function and

\begin{equation}
S(\omega ,\omega ^{\prime })=\sum_{mn}|g_{mn}|^2\delta (\omega -\Omega
_n)\delta (\omega ^{\prime }-\Omega _m)  \label{sww}
\end{equation}
is the so-called scattering function.

In the long time limit $\gamma (t)$ can, to a good approximation, be written as

\begin{equation}
\gamma (t)\cong \bar{\gamma}(T)\delta (t)
\label{gtT}
\end{equation}
and $\bar{\gamma}(T)$ is given by \cite{CyN}

\begin{equation}
\bar{\gamma}(T)=\frac 1{2\pi M_s}\int_0^\infty dE\ {\cal R}(E)\frac{\beta E\
e^{\beta E}}{(e^{\beta E}-1)^2}  \label{damp}
\end{equation}
where ${\cal R}(E)$ is the reflection coefficient of the ``potential'' $%
U^{\prime \prime }(\varphi _s)$ in the Schr\"{o}dinger-like equation (\ref
{Sle}). Notice that (\ref{damp}) is only valid if the states involved in (\ref{sww}) 
are scattering states (see section {\bf V.A}  below for details). 
One important point that should be emphasized here is that there are
parameters of the non-linear field equations for which the localized
solutions render $U^{\prime \prime }(\varphi _s)$ a reflectionless
potential. These are genuine solitons and for these the mobility is infinite.
One may realize this is what happens for the Bloch walls (\ref{fia10}) and (%
\ref{fia20}). In these cases, the ``potential'' appearing in (\ref{Sle}) can
be written as

\begin{equation}
U^{\prime \prime }(z)=\eta ^2(1-2%
\mathop{\rm sech}
\nolimits^2\eta z),
\label{u0}
\end{equation}
where $\eta ^2=A_1$ for vanishing anisotropy and $\eta ^2=2A_2$ for
vanishing external field. The spectrum of (\ref{u0}) contains a bound state
with zero energy

\begin{equation}
\psi _0=\sqrt{\frac \eta 2}%
\mathop{\rm sech}
(\eta z),\qquad
\kappa _0^2=0
\label{gm}
\end{equation}
which constitutes the translation mode of the domain wall (Goldstone mode),
and a continuum of quasiparticles modes (magnons) given \cite{MF} by
\begin{equation}
\psi_{n}(x)=\frac{1}{\sqrt{L}}\left[
\frac{k_{n}+i\eta \tanh(\eta z)}{k_{n}+i\eta}
\right]
e^{ik_{n}z},
\label{qp}
\end{equation}
where
\begin{equation}
k_{n}=\frac{2n\pi}{L}-\frac{\delta(k_{n})}{L}, \qquad
\delta(k)=\arctan \left[ \frac{2\eta k}{k^{2}-\eta^{2}} \right].
\end{equation}

It is known that the reflection coefficient ${\cal R}$ for a general
symmetric potential can be expressed in terms of the corresponding phase
shifts as \cite{CyN}

\begin{equation}
{\cal R}(k)=\sin ^2\left( \delta ^e(k)-\delta ^o(k)\right),  
\label{Rsin}
\end{equation}
where $\delta ^e$ and $\delta ^o$ are the even and odd scattering phase
shifts, respectively. Then, re-expressing (\ref{qp}) in terms of even 
and odd defined parity states, it is easy to prove that this potential  belongs
to the class of reflectionless because its phase shifts are given by
\begin{equation}
\delta ^{e,o}(k)=\arctan (\eta /k),
\label{ps0}
\end{equation}
that do not distinguish between odd and even parities.

Nevertheless, when both the anisotropy and external field are finite, the
reflection coefficient is nonvanishing and consequently the 2$\pi $%
-Bloch wall (\ref{figen}) has a finite mobility. In this case, the
spectral problem (\ref{Sle}) can be rewritten as

\begin{equation}
\left\{ -\frac{d^2}{dz^2}+V(z) \right\} \psi_n(z)=\kappa^2_n \psi_n(z),
\label{Sle2}
\end{equation}
where the potential $V(z)$ is expressed as
\begin{eqnarray}
V(z)&=&-\frac{
\mathop{\rm sech}
\nolimits^2(\rho)}{\lambda^2}
\left[\tanh(\frac{z}{\lambda}+\rho)\tanh(\frac{z}{\lambda}-\rho)-
\mathop{\rm sech}
\nolimits (\frac{z}{\lambda}+\rho)
\mathop{\rm sech}
\nolimits (\frac{z}{\lambda}-\rho)\right]
\nonumber \\
&&-\frac{\tanh^2(\rho)}{\lambda^2}
\left[\left(
\tanh(\frac{z}{\lambda}+\rho)\tanh(\frac{z}{\lambda}-\rho)-
\mathop{\rm sech}
\nolimits (\frac{z}{\lambda}+\rho)
\mathop{\rm sech}
\nolimits (\frac{z}{\lambda}-\rho)\right)^{2}
\right. \nonumber \\
&&-\left.
\mathop{\rm sech}
\nolimits ^2(\frac{z}{\lambda}+\rho)
\mathop{\rm sech}
\nolimits ^2(\frac{z}{\lambda}-\rho)
\left(\sinh(\frac{z}{\lambda}+\rho)\sinh(\frac{z}{\lambda}-\rho)
\right)^{2}
\right].
\label{PotV}
\end{eqnarray}  

Now, for all finite values of $\lambda $ and $\rho$, the translational invariance 
of the system persists and as a consequence the potential (\ref{PotV})
has a zero energy state that is given by
\begin{equation}
\psi _0\propto
\mathop{\rm sech}
(\frac{z}{\lambda}+\rho )+%
\mathop{\rm sech}
(\frac{z}{\lambda}-\rho )\ ,
\label{fi0g}
\end{equation}
which is nothing but the Goldstone mode of the Bloch walls for finite anisotropy
and external field.

In order to obtain an expression for the damping constant (\ref{damp}) we
need an expression for the odd and even phase shifts of (\ref{PotV}).
Unfortunately, their analytical evaluation is very complicated for all finite
values of $\lambda $ and $\rho $, and in what follows we study the situation
of large fields.

\section{2$\pi $-Bloch walls for large fields}

In this section we evaluate the scattering phase shifts in the situation of
large external fields and provide an explicit expression for the damping
constant.

\subsection{Scattering phase shifts}

In the case of large fields ($\rho \ll 1$) the Shr\"{o}dinger-like equation
(\ref{Sle2}) can be written as
\begin{equation}
\left\{
-\frac{d^2}{dz^2}+V(z) \right\} \psi_{n}(z)=(\kappa^{2}_{n}-\eta^{2}
-\frac{\rho^{2}}{\lambda^{2}})\psi_{n}(z),
\end{equation}
where the potential (\ref{PotV}) is now reduced to the sum of the
reflectionless contribution and a perturbation coming from the presence of the large
field , explicitly
\begin{equation}
V(z)=V_0(z)+\rho ^2V_1(z),  
\label{Vap}
\end{equation}
where
\begin{equation}
V_0(z)=-2%
\mathop{\rm sech}
\nolimits^2 \left( \frac{z}{\lambda} \right) 
\label{Vap0}
\end{equation}
and
\begin{equation}
V_1(z)=1-8\tanh ^2\left(\frac{z}{\lambda}\right)
\mathop{\rm sech}
\nolimits^2 \left( \frac{z}{\lambda} \right) .  
\label{Vap1}
\end{equation}

In order to obtain the even and odd scattering phase shifts $\Delta ^{e,o}$
corresponding to a particle in a one dimensional symmetric potential like (\ref{Vap}), 
we will use of a 1D version of the Fredholm theory \cite{Gott}, which states that
\begin{equation}
\pi A^{e,o}(E)\cot (\Delta ^{e,o})=1+{\cal P}\int_0^\infty dE\frac{%
A^{e,o}(E^{\prime })}{E-E^{\prime }},  
\label{piacot}
\end{equation}
where $\Delta ^{e,o}$ are the phase shifts originated by both contributions, the first
coming from the reflectionless potential $V_{0}$,
and the other associated to the high field perturbation $V_{1}$.
On the other hand the even and odd spectral functions, $A^{e,o}(E)$, can be calculated from the 
series expansion (see \cite{Gott} for details)
\begin{equation}
A(E)=-\langle E | V(z) | E \rangle +
{\cal P}\int_{0}^{\infty} \frac{dE_{1}}{E-E_{1}}
\left|
\begin{array}{cc}
\langle E |V(z) |E \rangle  &  \langle E|V(z)|E_{1} \rangle  \\
\langle E_{1}|V(z)|E\rangle & \langle E_{1}|V(z)|E_{1} \rangle \\
\end{array}
\right|+\cdots
\label{Aap}
\end{equation}
where ${\cal P}$ stands for the Cauchy principal value. Clearly the expression (\ref{Aap}) 
cannot be analytically evaluated to all orders. On the other hand, making use of (\ref{ps0})
and considering that $\rho$ is small enough, the expression for the the phase shifts (\ref{piacot})
can be written up to first order in $\rho^{2}$ as
\begin{equation}
\tan\Delta^{e,o}=\frac{\eta}{k}+
\rho^2 \frac{\pi A^{e,o}_{1}}
{1+2B^{e,o}_{0}},
\label{nova1}
\end{equation}
where
\begin{equation}
A_{1}=-\langle E |V_{1} (z)| E \rangle,
\label{yyy}
\end{equation}
and
\begin{equation}
B^{e,o}_{0}={\cal P}
\int_{-\infty}^{\infty}
\frac{A^{e,o}_{0}(k^{\prime}) k^{\prime} dk^{\prime}}{k^2-k^{\prime 2}}, \qquad
A_{0}=-\langle E|V_{0}(z)|E\rangle.
\label{rmi}
\end{equation}

Using a convenient basis set, the three expressions given by (\ref{yyy}) and (\ref{rmi}) can be 
analytically evaluated (see the apendix) yielding
\begin{equation}
A^{e,o}_{0}(k)=\frac{2M}{\hbar^2}\left[
\frac {1}{\pi k \lambda } \pm \frac{1}{\sinh (\pi k \lambda)}
\right],  
\label{aeok2}
\end{equation}
\begin{equation}
A^{e,o}_{1}(k)=\frac{8 \rho^2 M}{\hbar^2}\left[
\frac {1}{\pi k \lambda } \mp  (2k^2-\lambda^{-2})\frac{\lambda^2}{\sinh (\pi k \lambda)}
\right],  
\label{aeok}
\end{equation}
\begin{equation}
B^{e,o}= \pm \frac{4M}{\hbar^2} 
\sum _{n=1}^\infty (-1)^{n+1} 
\frac {n}{((k/\lambda)^{2} + n^2)}.
\label{beo}
\end{equation}
Now we can finally write down an expression for the phase shifts $\Delta^{e,o}$ by substituting 
(\ref{aeok}) and (\ref{beo}) in (\ref{nova1}). In so doing one gets
\begin{equation} 
\tan \Delta ^{e,o}(k) =  
\frac{{\textstyle \eta}}{{\textstyle k}}+
\frac{ {\textstyle 8 \pi \hbar^{-2} M \rho^2 }}{ {\textstyle 1} \pm 
\frac{{\textstyle 8M}}{{\textstyle \hbar^2}} 
{\displaystyle \sum _{n=1}^\infty} {\textstyle (-1)^{n+1}} 
\frac {{\textstyle n}}{{\textstyle (k/\lambda)^{2} + n^2}}}
\left[
\frac{{\textstyle 1}}{{\textstyle k \pi \lambda}} \mp 
\frac{{\textstyle 2k^2 \lambda^{2}-1}}{{\textstyle 3 \sinh(k \pi \lambda)}}
\right]
\label{deltaeo}
\end{equation}

Looking at (51) we realize that, whereas $\Delta ^{e}(k)$ remains almost unchanged as
a function of $k$ for $\rho \neq 0$ (see Fig.1), $\Delta ^{o}(k)$ presents a completely 
different structure. As it is shown in Fig.2, the odd phase shift for $\rho \neq 0$ 
display a $\pi$ amplitude discontinuity characteristic of a resonance. This discontinuity
becomes sharper as $\rho \rightarrow 0$, giving no contribution to the reflection 
coefficient, because as expected from (\ref{ps0}), the even and the odd phase shifts approach
each other for small $\rho$ (see Fig.3). In this situation the $\pi$ amplitude discontinuity
is still present, but as an isolated one, does not contribute to the reflection coefficient.
We can now turn our atention to the behaviour of the phase shifts near $k=0$. In order to do
this we will use the 1D\ version of the Levinson's theorem \cite{Bart} for one-dimensional 
symmetric potentials. Levinson's theorem establishes that:
\begin{eqnarray}
\Delta ^e(k &=&0)=\pi (n^e-\frac 12), \nonumber \\
\Delta ^o(k &=&0)=\pi n^o,
\label{levi}
\end{eqnarray}
where $n^e$ and $n^o$ are the number of even and odd parity bound states. 
As it can be seen in Fig. 1, the even parity state phase shift is always $\pi/2$ for $k=0$ for 
any finite value of $\rho $, then from (\ref{levi}) it is clear that our system has an even parity
bound state for any value of $\rho$. The existence of this state is in complete agreement with
the translational invariance of the system and corresponds to the Goldstone mode. The analysis of
the existence of an odd parity bound state is slightly different. As can be seen for (\ref{ps0})
the correct definition of $\delta^o$ according to (\ref{levi}) is
\begin{equation}
\delta^{o}=
\left\{
\begin{array}{ll}
0, & \mbox{ $ k=0 $ } \\
\arctan (\eta / k), & \mbox{$k \neq 0$}
\end{array}
\right.
\label{pinky}  
\end{equation}
because there is no odd parity bound state for the $\rho=0$ case. Following the same idea
and realizing that (\ref{deltaeo}) behaves in the same way as (\ref{ps0}) for small enough 
$k$, we conclude that $\Delta^o(0)=0$ and therefore there is no odd bound state in the 
$\rho \neq 0$ case. Actually, $\Delta^e(0) \approx \Delta^o(0)$ for large values of k
whereas they start to deviate from one another as k decreases. Equation (\ref{pinky}) 
represents an extreme situation when $\delta^o(k)$ discontinously jumps from $\delta^e(k)$
to $\delta^o(k)=0$ when $k \rightarrow 0$. This is the only way we can reconcile the absence of an
odd bound state and the reflectiolessness of the potential.
Therefore, the spectrum of (\ref{Vap}) is composed by: 
i) the $\psi _0$ solution (\ref{fi0g}) corresponding to the
translation mode of the wall (Goldstone mode) and ii) the $\psi _k$ solutions
which constitute the continuum modes and correspond to magnons.

\subsection{The damping coefficient}

In order to find the damping coefficient we must compute ${\cal R}(k)$. This can be done by 
inserting (\ref{deltaeo}) into the general expression
\begin{equation}
{\cal R}(k)=\sin^{2}(\Delta ^{e}(k)-\Delta ^{o}(k)).
\end{equation}
Although we have its analytical form in the present approximation, we had better plot it 
for the whole range of the momentum $ k$ for various ratios of $ A_{2}/A_{1}$ 
(anisotropy/external field) (see Fig. 4).
Having done that, one can immediately integrate this function in expression (\ref{damp}) 
which finally allows us to describe the damping as a function of the temperature as shown 
in Fig.5. As it can be seen, the damping constant is linear for the whole range of temperatures. 
This result can be obtained directly from (\ref{damp}). For $T$ high enough the
damping constant can be approximated by
\begin{equation}
\bar{\gamma}(T)\simeq \frac 1{2\pi M_s\beta }\int_0^\infty dE\ \frac{{\cal R}%
(E)}E\propto T  
\label{g1}
\end{equation}
which is linear on $T$, independently of the explicit form of ${\cal R}(E)$. In the 
low temperature regime we can write
\begin{equation}
\bar{\gamma}(T)\simeq \frac 1{2\pi M_s}\int_0^\infty dE\ {\cal R}(E)\beta
Ee^{-\beta E} 
\label{aoc}
\end{equation}
where $E$ always presents a gap given by $E^2_{g}=[(\eta\lambda)^2+\rho^2]\hbar^2c^2/\lambda^2$ 
and therefore and exponentially small damping. However, this expression must be studied 
in the limit in which we are interested; namely, $\rho \rightarrow 0$. Unlike $\bar{\gamma}(T)$ 
given by (\ref{g1}) the expression (\ref{aoc}) cannot be analitically estimated in a trivial way 
because ${\cal R}(E)$ has only been computed numerically. Here it must be stressed that this result 
is only reliable for not too low temperatures because we have employed the odd phase shift 
$\Delta^o(k)$ from Fig.2 and its computation clearly do not account for its correct values for 
$k \rightarrow 0$. This behaviour is due to the fact that for very low energies  one can not 
approximate $A(E)$ in (\ref{Aap}) by its first term although this approximation works well for the even 
phase shift $\Delta^e(k)$. This turns out to be a good description of the mobility of the wall 
for extremely high fields if we keep the above explanation in mind. 

\section{Conclusions}

In this paper we have analyzed the dissipative dynamics of Bloch walls in a one dimensional 
anisotropic ferromagnet in the presence of an external magnetic field. In particular, we have 
considered the limit of high magnetic fields although there is no reason why one should not 
apply the same methods to the low field case. The only difference is that the scattering 
problem with which one has to deal is more straightforward in the high field case.
Our predictions are that the damping coefficient $\bar{\gamma}(T)$ presents a linear behaviour 
as is plotted in Fig.5, for a vast range of temperatures.

\section{Acknowledgments}

One of us (M.A.D.) would like to acknowledge the financial support from FAEP
(Fundo de Apoio ao Ensino e Pesquisa)\ during his visit to UNICAMP. A.V.F. wishes 
to thank FAPESP (Funda\c{c}\~{a}o de Amparo \`{a} Pesquisa no Estado de S\~{a}o Paulo) for a 
scholarship, A.O.C. kindly acknowledges the partial support from the CNPq (Conselho
Nacional de Desenvolvimento Cientif\'{\i}co e Tecnol\'{o}gico) whereas A.H.C.N. acknowledges
support from the A.P. Sloan foundation and support provided by the DOE for research at Los
Alamos National Laboratory.

\section{Appendix}

In this appendix we show how to obtain the expressions of the even and odd
spectral functions (\ref{rmi}) and (\ref{yyy}). Suppose we have a particle in a one 
dimensional symmetric potential of the form $V=V_0+gV_1$ confined to a region ($-L,+L)$ with 
$L$ much larger than the range of the potential $V$. The asymptotic form of the
wave functions for $V \neq 0$ are given by
\begin{eqnarray}
|z\rangle ^e &=&\sqrt{\frac 1L}\cos (k|z|+\Delta ^e(k)), \nonumber \\
|z\rangle ^o &=&\sqrt{\frac 1L}%
\mathop{\rm sgn}
(z)\sin (k|z|+\Delta ^o(k)),
\label{figa}
\end{eqnarray}
for $|z|\rightarrow \infty$. If $V=0$ the wave 
functions have the same structure as in (\ref{figa}) with $\Delta ^{e,o}=0$.
Because the wave functions must vanish at $z=\pm L$ one realizes that
\[
\frac{\delta E_n}{\Delta E_n}=-\frac 1\pi \Delta ^{e,o}
\]
where $\Delta E_n=E_{n+1}^0-E_n^0$ and $\delta E_n=E_n-E_n^0.$ Following
closely the prescriptions given in \cite{Gott} for the 1-D case, the spectral 
functions (\ref{rmi}) and (\ref{yyy}) are given by
\begin{equation}
A^{e,o}_{i}(E)=-\int_{-\infty }^{+\infty }dzV_i(z)\langle z|E\rangle
_{e,o}\langle E|z\rangle _{e,o}, \qquad i=0,1 
\label{Aap2}
\end{equation}
where the states $\langle E|z\rangle_{e,o}$ can be obtained by taking
\begin{equation}
\langle z|E\rangle_{e,o}=\lim_{L \rightarrow \infty}\frac{|z \rangle}{\sqrt{\Delta E_{n}}}.
\end{equation}
Explicitly,
\begin{equation}
\langle z|E\rangle =\frac 1{\sqrt{2\pi k}}
\left\{
\begin{array}{ll}
\cos (kx), & \mbox{ for even parity} \\
\sin (kx), & \mbox{for odd parity}
\end{array}
\right.  
\label{zeo}
\end{equation}

Inserting (\ref{zeo}) and (\ref{Vap1}) in (\ref{Aap2}) we have

\begin{equation}
A_{1}^{e}+A_{1}^{o}=\frac{8M\rho ^2}{\pi \hbar^2 k} 
\int_{-\infty }^{+\infty}
\mathop{\rm sech}
\nolimits ^2(\frac{z}{\lambda}) 
\tanh ^2 (\frac{z}{\lambda})dz,
\end{equation}

which can be easily evaluated with the substitution $y=\tanh z/\lambda$, yielding

\begin{equation}
A_{1}^{e}+A_{1}^{o}=\frac {16 \rho ^2 M}{ \pi \hbar^2 \lambda k}.  
\label{amas}
\end{equation}

On the other hand we have

\begin{equation}
A_{1}^{e}-A_{1}^{o}=\frac{8M\rho ^2}{\pi \hbar^2 k} 
\int_{-\infty }^{+\infty }
\mathop{\rm sech}
\nolimits ^2(\frac{z}{\lambda}) 
\tanh ^2 (\frac{z}{\lambda}) \cos(2k \lambda)dz,
\end{equation}
that can be writen as

\begin{equation}
A_{1}^{e}-A_{1}^{o}=\frac{8M\rho ^2}{\pi \hbar^2 k} 
\int_{-\infty }^{+\infty } \left(
\mathop{\rm sech}
\nolimits ^2(\frac{z}{\lambda})-
\mathop{\rm sech}
\nolimits ^4(\frac{z}{\lambda}) \right)
\cos(2k \lambda)dz
\end{equation}
which can be analytically evaluated \cite{Grad} yielding

\begin{equation}
A_{1}^{e}-A_{1}^{o}=\frac {16 \rho ^2 M}{ \hbar^2 \sinh (\pi k \lambda)}
\left[\frac{1}{3}-\frac{2 k^2 \lambda^2}{3}
\right].
\label{amen}
\end{equation}

Therefore combining (\ref{amas}) and (\ref{amen}) we have (\ref{aeok}). In
the same fashion it can be shown that 
\begin{equation}
A_{0}^{e}+A_{0}^{o}=\frac {4 M}{ \pi \hbar^2 k \lambda} \qquad 
$and$ \qquad
A_{0}^{e}-A_{0}^{o}=\frac {{\textstyle 4 M}}{ {\textstyle \hbar^2 \sinh ( \pi k \lambda)}}
\end{equation}
which immediately gives (\ref{aeok2}). Now we can evaluate the Cauchy principal value 
in (\ref{rmi}) which reads
\begin{equation}
B_{0}^{e,o}=\frac{2M}{\hbar^2}{\cal P}\int_{-\infty}^{+\infty}
\left[\frac{1}{\pi k^{\prime} \lambda} \pm \frac{1}{\sinh (\pi k^{\prime} \lambda)}
\right]
\frac{k^{\prime} dk^{\prime}}{k^2-k^{\prime 2}}.
\label{bras}
\end{equation} 
The first term on the right hand side of (\ref{bras}) is clearly zero. Therefore, using the
product expansion of $\sinh (\pi z)$ fuction \cite{Grad} its second term becomes
\begin{equation}
B_{0}^{e,o}=\pm \frac{2M}{\pi \hbar^2}{\cal P}\int_{-\infty}^{+\infty}\frac{dq}{k^2-q^2}
\prod_{n=1}^{\infty}\frac{n^2}{n^2+q^2},
\end{equation}
where $q=k/\lambda$. Going to the complex plane, the previous expression can be analitically 
evaluated as
\begin{equation}
B_{0}^{e,o}=\pm \frac{4M}{\hbar^2}\sum_{n=1}^{\infty}\frac{(-1)^n n}{(k/\lambda)^2+n^2}.
\end{equation}

\begin{figure}
\vbox to 9.0cm {\vss\hbox to 9cm
 {\hss\
   {\includegraphics{eve.ps}
  }
  \hss}
 }
\caption{The even phase shift as a function of the momentum for three different situations.
The continuous line corresponds to $\rho=0.14$, the dotted line to $\rho=0.31$ and the dashed line 
to $\rho=0.60$.}
\label{abs}
\end{figure}
\begin{figure}
\vbox to 9.0cm {\vss\hbox to 9cm
 {\hss\
   {\includegraphics{odd.ps}
  }
  \hss}
 }
\caption{The odd phase shift as a function of the momentum. The triangles correspond to
$\rho=0.14$ and the circles to $\rho=0.31$.}
\label{abs}
\end{figure}
\begin{figure}
\vbox to 9.0cm {\vss\hbox to 9cm
 {\hss\
   {\includegraphics{phase.ps}
  }
  \hss}
 }
\caption{The even and odd phase shifts when $\rho=0.14$. As can be seen they approch each 
other as the ratio $A_1/A_2$ increases, thats means $\rho \rightarrow 0$, and the only difference 
comes from the singular point in the odd phase shift contribution.} 
\end{figure}
\begin{figure}
\vbox to 9.0cm {\vss\hbox to 9cm
 {\hss\
   {\includegraphics{reflex.ps}
  }
  \hss}
 }
\caption{The reflection coefficient as a function of the momentum. The continuous line for
$\rho=0.14$, the dotted line for $\rho=0.31$ and the dashed line for $\rho=0.60$.} 
\end{figure} 
\begin{figure}
\vbox to 9.0cm {\vss\hbox to 9cm
 {\hss\
   {\includegraphics{findam.ps}
  }
  \hss}
 }
\caption{The damping coefficient as a function of the temperature for different 
values of $\rho$. The continuous line for $\rho=0.14$, the dashed line for $\rho=0.31$ and the
dotted line for $\rho=0.60$.} 
\end{figure}
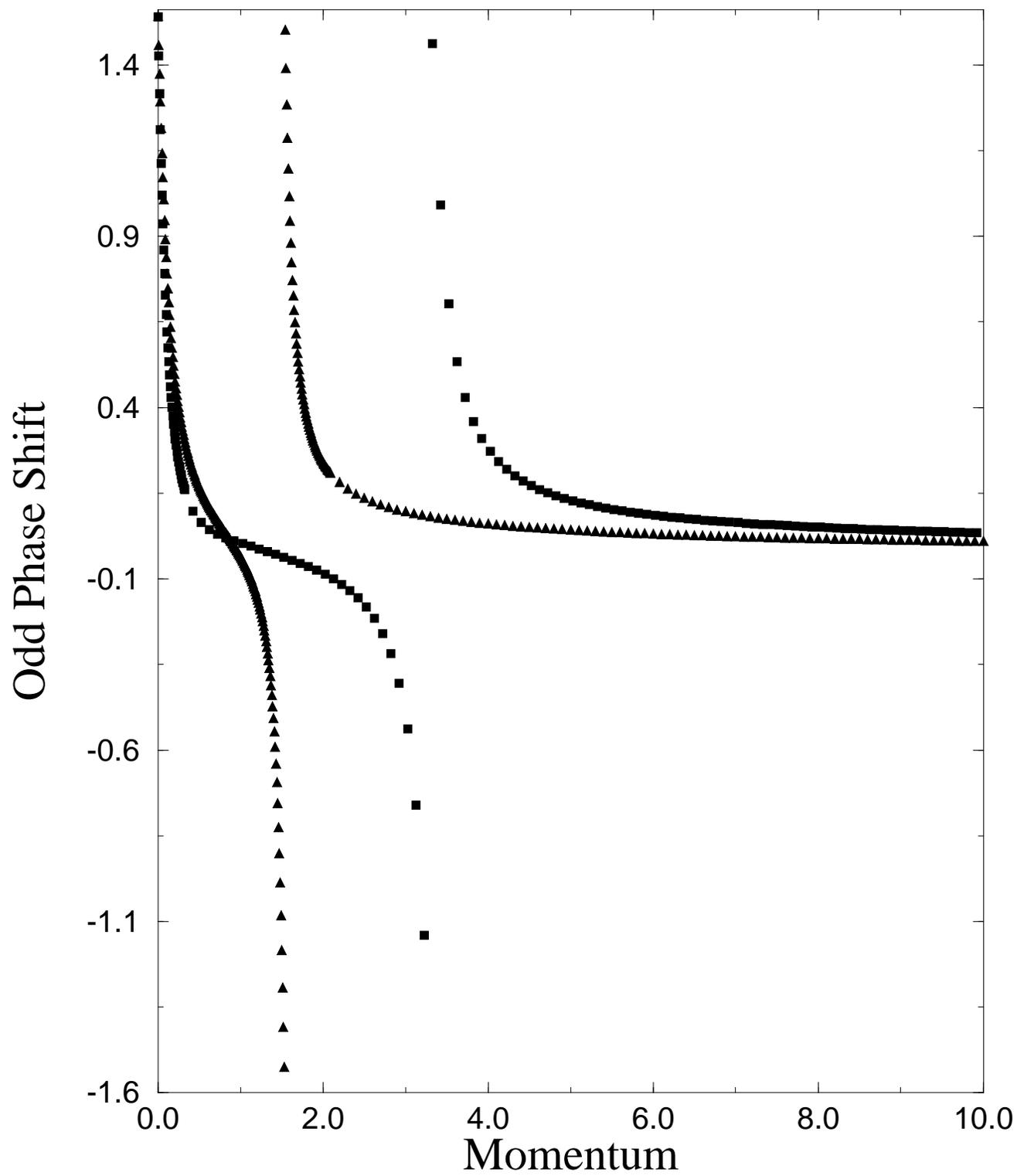
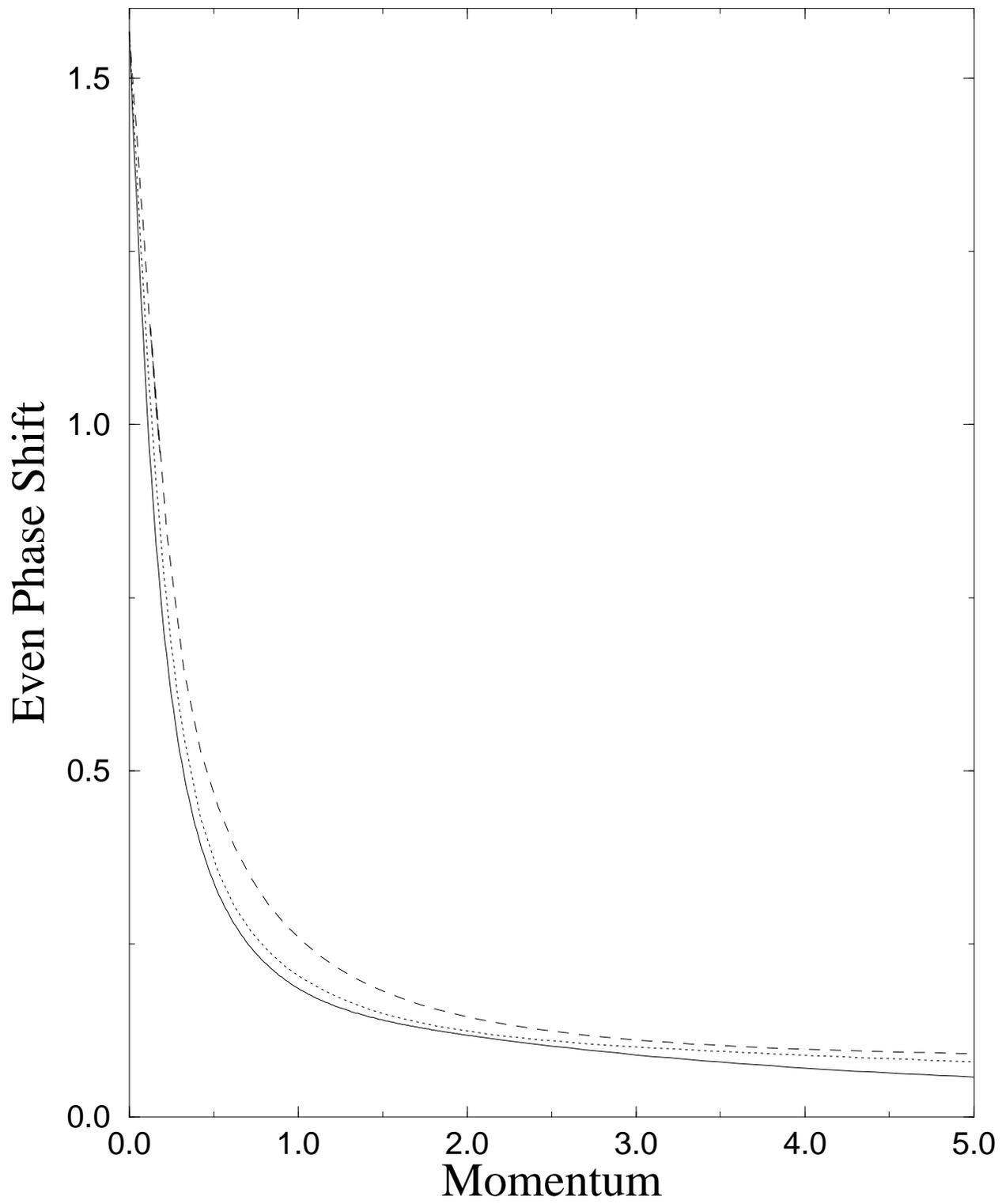
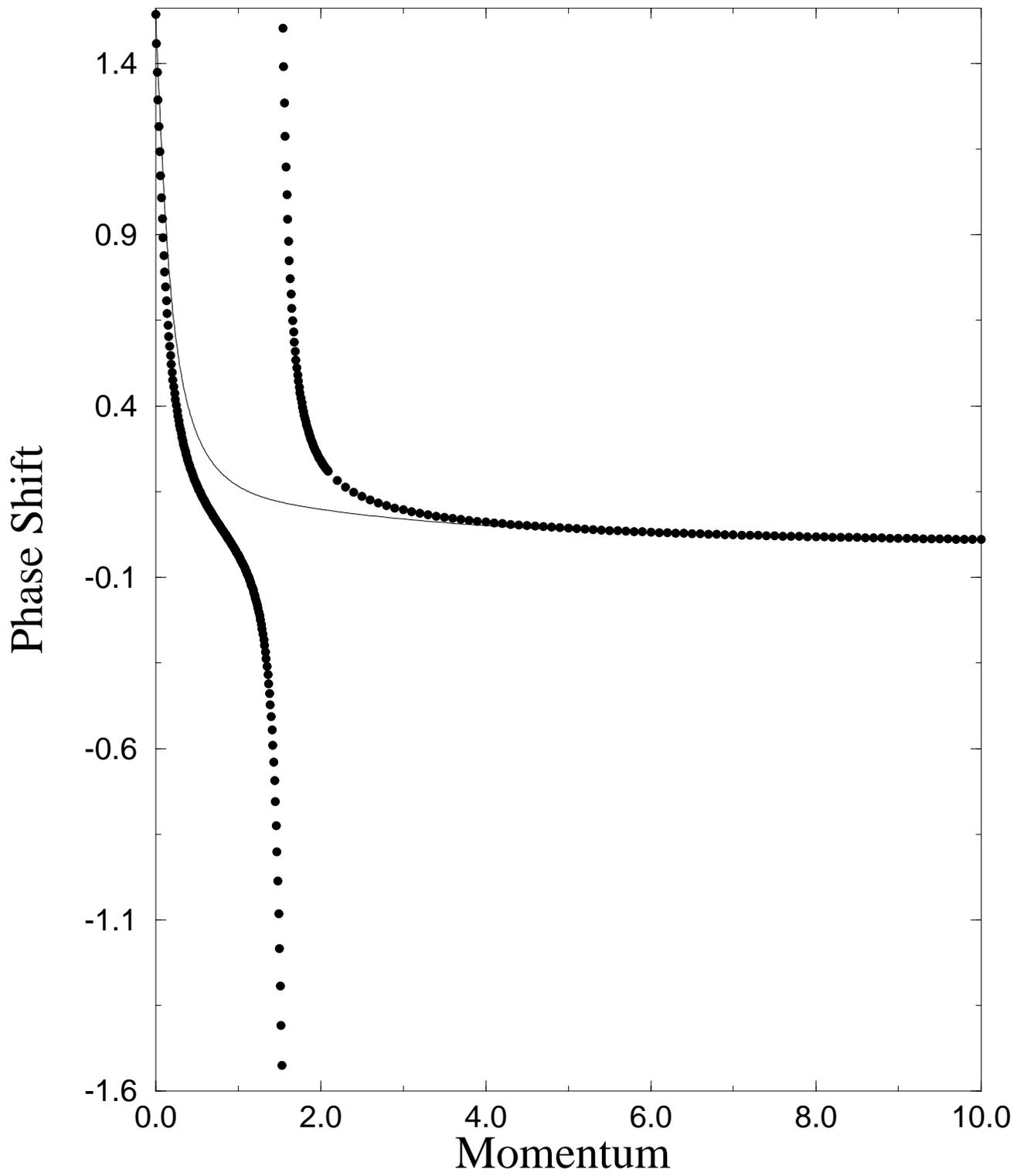
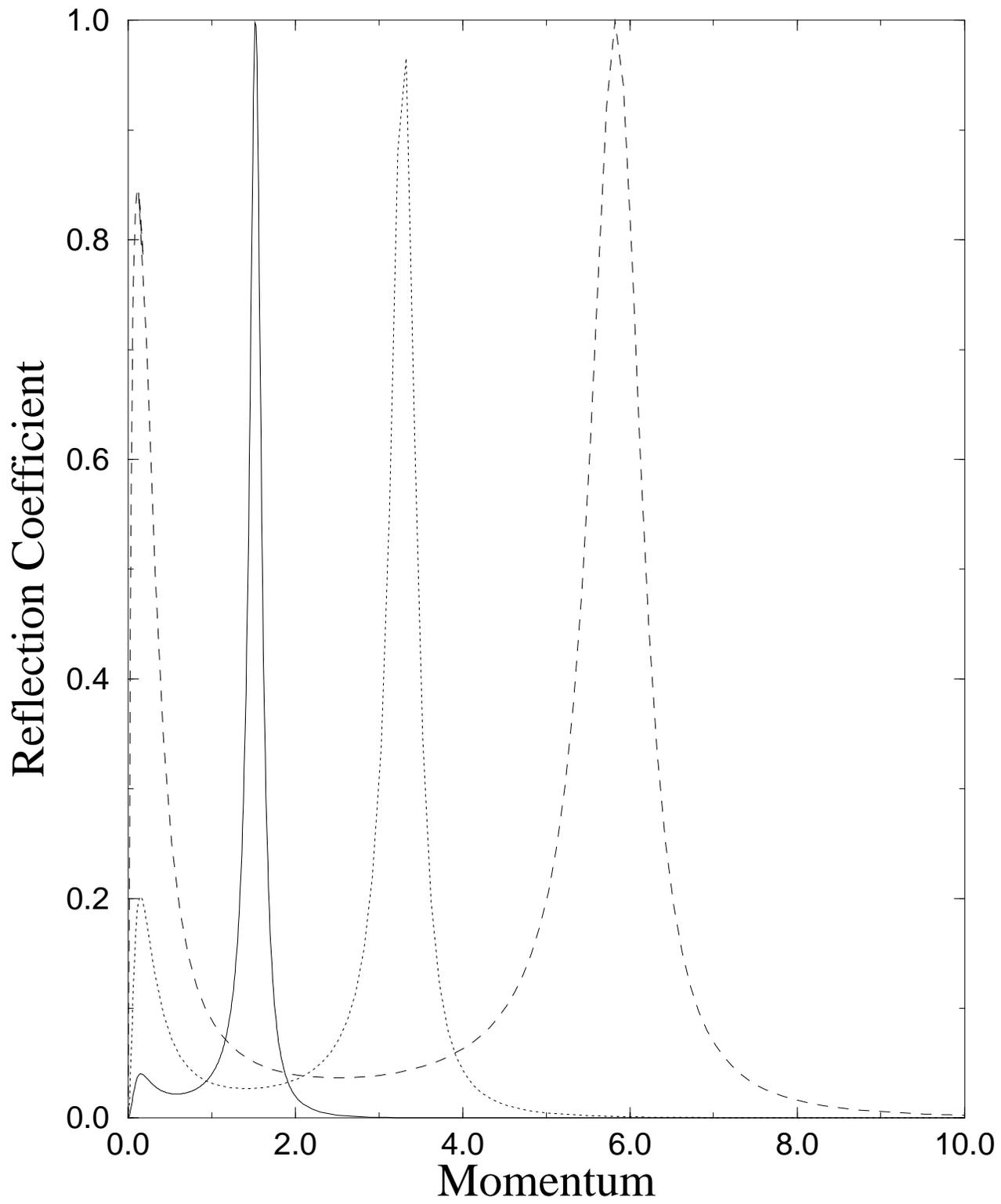
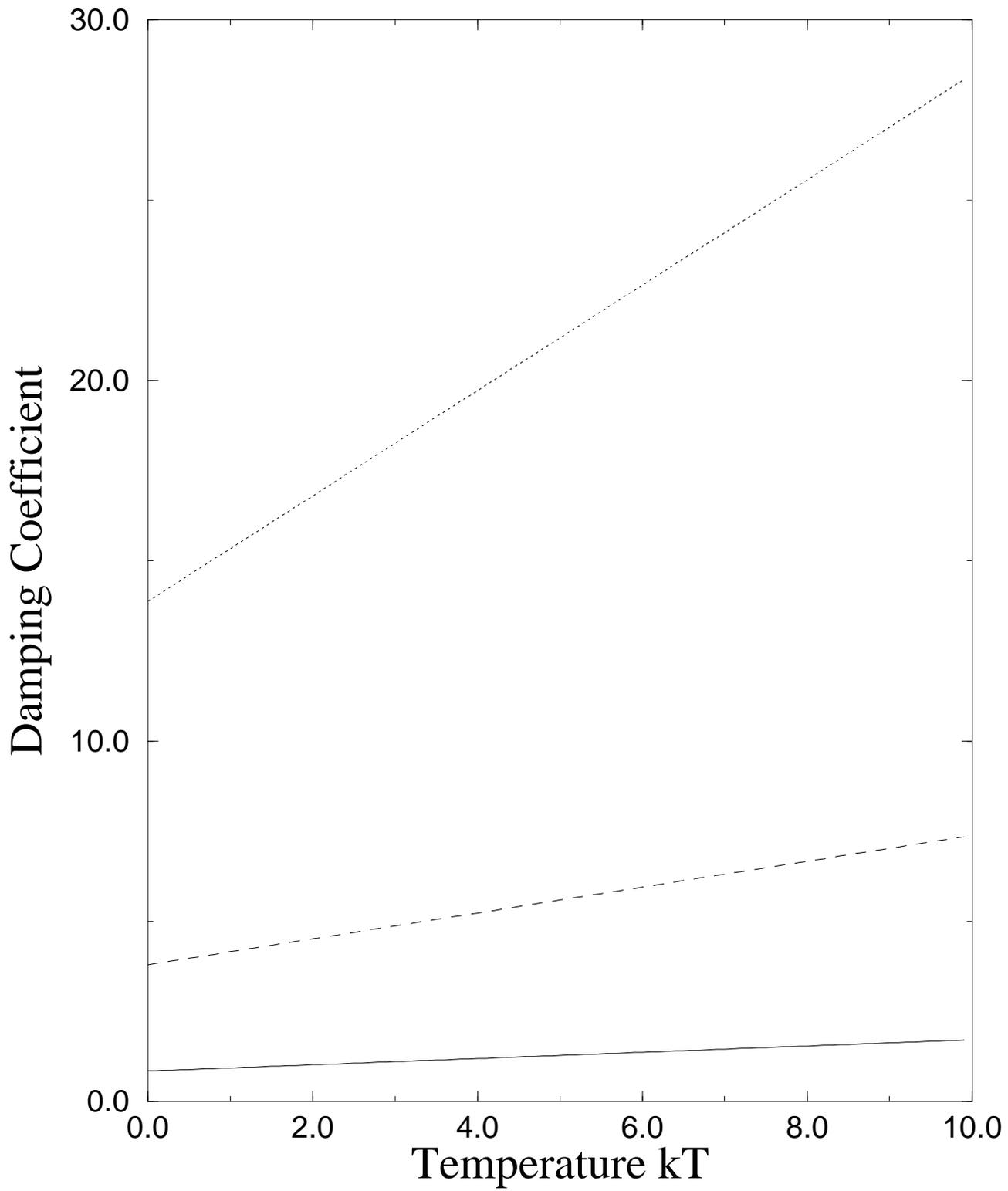
\end{document}